\documentclass[12pt]{article}
\usepackage{graphicx,amsmath,amsfonts,amssymb,fancybox}% Include figure files and higher math
\usepackage{dcolumn}% Align table columns on decimal point
\usepackage{bm}% bold math
\usepackage{color}
\usepackage{subfigure}
\usepackage{array}
\usepackage{multirow}
\begin{document}

\title{Surface Effects on the Piezoelectricity of ZnO Nanowires}

\author{Shuangxing Dai,Harold S. Park\footnote{Electronic address: parkhs@bu.edu}\\
  {Department of Mechanical Engineering, Boston University, Boston, MA 02215}}

\date{\today}
\maketitle

\begin{abstract}

We utilize classical molecular dynamics to study surface effects on the piezoelectric properties of ZnO nanowires as calculated under uniaxial loading.  An important point to our work is that we have utilized two types of surface treatments, those of charge compensation and surface passivation, to eliminate the polarization divergence that otherwise occurs due to the polar (0001) surfaces of ZnO.  In doing so, we find that if appropriate surface treatments are utilized, the elastic modulus and the piezoelectric properties for ZnO nanowires having a variety of axial and surface orientations are all reduced as compared to the bulk value as a result of polarization-reduction in the polar [0001] direction.  The reduction in effective piezoelectric constant is found to be independent of the expansion or contraction of the polar (0001) surface in response to surface stresses.  Instead, the surface polarization and thus effective piezoelectric constant is substantially reduced due to a reduction in the bond length of the Zn-O dimer closest to the polar (0001) surface.  Furthermore, depending on the nanowire axial orientation, we find in the absence of surface treatment that the piezoelectric properties of ZnO are either effectively lost due to unphysical transformations from the wurtzite to non-piezoelectric d-BCT phases, or also become smaller with decreasing nanowire size.  The overall implication of this study is that if enhancement of the piezoelectric properties of ZnO is desired, then continued miniaturization of square or nearly square cross section ZnO wires to the nanometer scale is not likely to achieve this result.

\end{abstract}

% \pacs{}

\section{Introduction} 

Piezoelectricity has long been a property of interest for bulk materials as it enables the direct conversion of mechanical strain into harvestable electrical energy~\cite{roundyJIMSS2005,antonSMS2007}.  While the interest in bulk piezoelectric materials has existed for some time, there has recently been significant interest in studying the piezoelectric behavior and properties of nanomaterials~\cite{voonJAP2011}.  Much of the interest has centered around ZnO, which was recently utilized by Wang \emph{et al.}~\cite{wangSCIENCE2006,songNL2006} to generate electrical energy through application of bending deformation via an atomic force microscope (AFM).  ZnO has proven to be a versatile choice for nanoscale piezoelectrics as it exhibits both semiconducting and piezoelectric properties~\cite{wangSCIENCE2006}, because it can be fabricated in a wide range of nanometer shapes and geometries~\cite{wangAFM2004}, and because it has the largest piezoelectric response of any tetrahedrally bonded semiconductor~\cite{dalcorsoPRB1994}.  Since the initial discovery in 2006, there have since emerged, though not without controversy~\cite{alexeAM2008}, a wide range of interesting applications involving ZnO~\cite{wangBOOK2011,xuNN2010}, GaN~\cite{suAPL2007}, and other nanostructures~\cite{Cha:2010ik,Xu:2010cs,Zhu:2010kh,Voon:2011wb,Sun:2010cm}.

In addition to the wide range of potential applications, recent experimental~\cite{zhaoNL2004} and computational~\cite{xiangAPL2006,mitJPCC2009,daiJAP2011,momeniAM2012} work has suggested that due to nanoscale surface effects, ZnO nanostructures may exhibit different piezoelectric properties than bulk ZnO.  These non-bulk piezoelectric properties may couple with the recent finding that ZnO nanostructures exhibit mechanical properties, and specifically Young's modulus that also shows a clear size-dependence due to surface effects~\cite{chenPRL2006,agrawalNL2008,parkMRS2009} to potentially enable ZnO nanowires (NWs) to produce more mechanical strain energy that can be converted through the piezoelectric effect into harvestable electrical energy than bulk ZnO.  

However, a key issue that has not been resolved is how surface effects impact the piezoelectric properties of ZnO NWs.  In other words, will making ZnO NWs smaller lead to enhanced piezoelectric properties?  We note that the NW geometry has been studied for other materials, for example using molecular dynamics (MD) for BTO~\cite{zhangNANO2009,zhangNANO2010,zhangJAP2011}, for GaN NWs using \emph{ab initio} techniques~\cite{agrawalNL2011}, and also using recently developed analytical theories~\cite{shenJMPS2010,yanJPD2011,morozovskaPRB2010,majidiJMPS2010,majdoubPRB2008}.  The piezoelectric properties of ZnO nanostructures, though excluding surface effects, have also been studied primarily using \emph{ab initio} calculations~\cite{karanthPRB2005,alahmedPRB2008}; the surface piezoelectric properties of ZnO were recently studied by~\cite{daiJAP2011}, though the effects on one-dimensional NWs were not considered.  A recent MD study did consider ZnO nanobelts~\cite{momeniAM2012}, though only for the [0001] orientation in which the transverse surfaces are not the polar (0001) surfaces and in which surface treatment, as described in the following paragraph, were not considered.  The one-dimensional NW geometry is critical to study and understand because it is most often utilized in application~\cite{wangBOOK2011}, where the NWs are subject to axial~\cite{Cha:2010ik,Yang:2009hf,Xu:2010cs}, bending~\cite{wangSCIENCE2006,Wang:2007jo} or shear deformations~\cite{Majidi:2010ib}.  Furthermore, ZnO NWs can be synthesized with a variety of axial and surface orientations~\cite{zhaoNL2004}, and cross sectional geometries~\cite{agrawalNL2011}, which will impact the piezoelectric properties in different fashions.  Therefore, a comprehensive understanding of how surface effects impact the piezoelectric properties of ZnO NWs, and how the piezoelectric properties of ZnO NWs vary with different surface and axial orientations of ZnO NWs remains unresolved.  It is the purpose of this work to shed insight into these issues, by virtue of classical MD simulations.

A related, and important issue this work addresses is the effect of the treatment of the polar ZnO  $(0001)$ surfaces on the piezoelectric properties.  Specifically, as previously discussed by~\cite{taskerJPC1979} and subsequently by other researchers~\cite{nogueraJPCM2000,Wander:2001vr,Kresse:2003to}, when there is a dipole moment in the repeat unit normal to the surface of an ionic crystal, the electrostatic energy diverges, and the surface energy goes to infinity.  Because of this, there are typically three techniques that are employed in atomistic simulations to eliminate this effect: charge compensation (CC)~\cite{Kresse:2003to,daiJAP2011}, surface reconstruction (SR)~\cite{Meskine:2011cd,Du:2008tx} and surface passivation (SP) or adsorption~\cite{Stengel:2011gj}.  These stabilization techniques are utilized because such reconstructions and passivation have been observed experimentally~\cite{Lauritsen:2011db,Lai:2010gw,Dulub:2003kx}, and they have also been widely used in first principles calculations~\cite{Dag:2011hg,Wander:2001wh}.  In contrast, they have rarely been utilized in classical MD simulations~\cite{Jia:2007hp,daiJAP2011} to avoid the divergence of the electrostatic potential.  While it is crucial to adopt one of these surface treatments for electrostatic stabilization, such treatments have not been utilized in previous MD studies of the size-dependent elastic properties of ZnO~\cite{kulkarniNANO2005}, or \emph{ab initio} studies of the piezoelectric properties of other ZnO NWs~\cite{Agrawal:2010go,agrawalNL2011}.  We will demonstrate the issues that arise in the electromechanical properties of ZnO if no surface treatment is undertaken.

\section{Methods}

We utilized classical MD to study the piezoelectric properties of ZnO NWs.  Specifically, we used the open source GROMACS 4.0 molecular simulation code~\cite{hessJCTC2008} while employing the Buckingham potential of~\cite{binksSSC1994} to model the various Zn-O interactions.  The Binks potential has been widely utilized to study the mechanical deformation of ZnO NWs~\cite{kulkarniNANO2005}.  However, until recent work by the authors for both bulk ZnO~\cite{daiNANO2010}, and subsequently for the surfaces of ZnO~\cite{daiJAP2011}, the performance of the Binks potential for the piezoelectric properties of ZnO had not been investigated.  Both works found the accuracy of the classical Binks potential to be comparable to benchmark \emph{ab initio} calculation results~\cite{dalcorsoPRB1994}.  

The lattice parameters we used for ZnO were $a_{0}$=3.2709 $\AA$, $c_{0}$=5.1386 $\AA$ and $u$=0.3882.  For the electrostatic interactions, we utilized the approach of~\cite{fennellJCP2006}, who improved on the original work of~\cite{wolfJCP1999} by ensuring that the electrostatic potential and force smoothly truncate at the cut-off radius, which results in stability for MD simulations~\cite{Fukuda:2008ux}.  The approach of~\cite{fennellJCP2006}, which enables the convergent calculation of the electrostatic energies and forces using a finite cut-off distance, is needed for the present simulations due to the fact that the standard Ewald method assumes an infinite, periodic crystal which is certainly not the case here due to the surface-dominated NW geometries.  The errors introduced by using the Ewald summation as compared to the Wolf technique for finite-sized NWs were recently quantified by~\cite{gdoutosIJNME2010}.  The specific parameters for the Fennell method that we utilized for ZnO were $\alpha=3$ nm$^{-1}$, $r_{c}$=1~nm; these parameters were previously found to give convergent results for the piezoelectric properties of ZnO~\cite{daiJAP2011}.

\begin{figure}
\centering
\subfigure[]{
\includegraphics[width=2.0in]{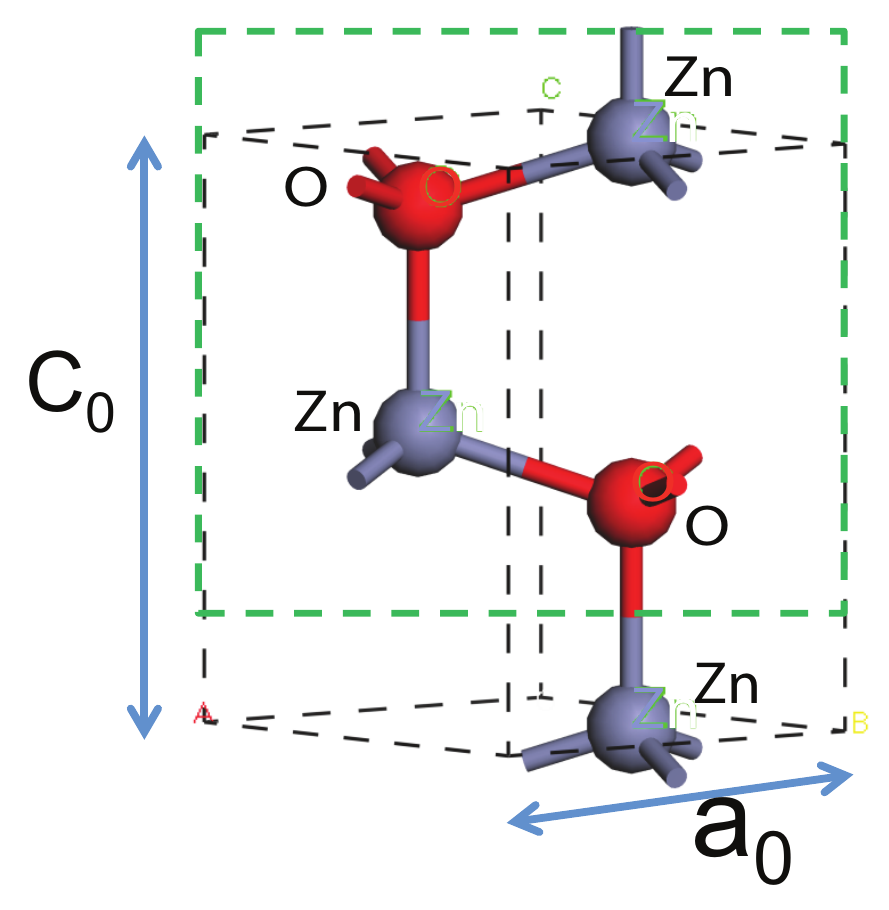}
\label{subfig:uco}
}
\subfigure[]{
\includegraphics[width=3.3in]{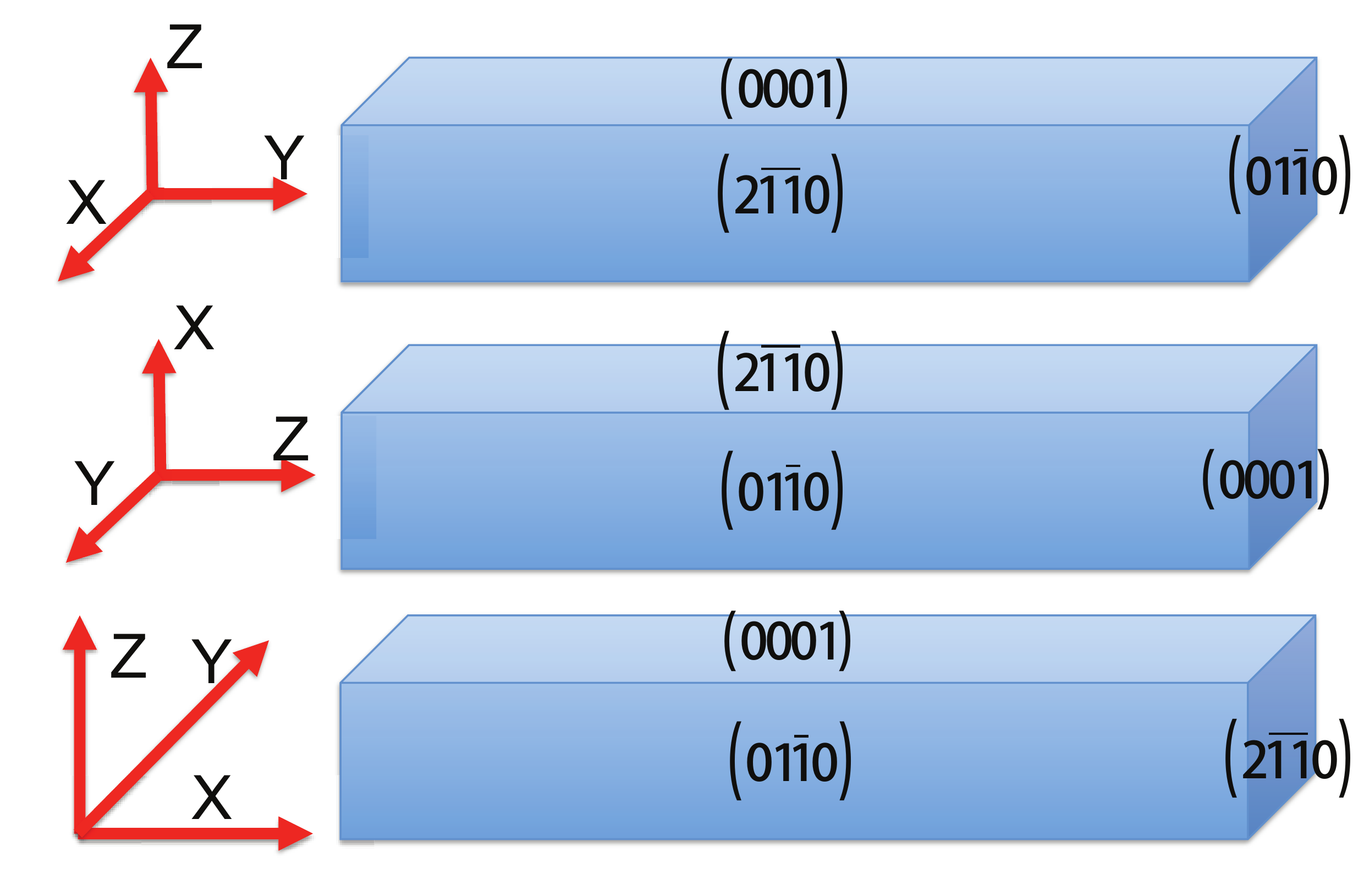}
\label{subfig:NW3}
}
\caption{(a) Four atom unit cell (in green rectangular box) for the wurtzite crystal structure; (b) Three ZnO NWs considered in this work.  From top, the axial orientations are along the $[01\overline{1}0]$, $[0001]$ and $[2\overline{11}0]$ directions.  The $z$-direction is chosen to be along the $[0001]$ direction for all NW orientations.}
\label{ucori}
\end{figure}

We considered nearly square cross section ZnO NWs with cross sectional lengths ranging from 2 to 4 nm.   We did not consider NWs with cross sectional sizes smaller than 2 nm because at these small sizes, a transformation into either a nonpiezoelectric d-BCT lattice structure~\cite{Kulkarni:2008kp, Sarasamak:2008vh, Wang:2008ki, Wang:2007gd} or a shell structure~\cite{kulkarniNANO2005} occurred as was previously predicted using MD simulations.  The specific combinations of axial and surface orientations we considered are illustrated in Fig. \ref{ucori}, with the NW sizes summarized in Table \ref{tab:size}, where the $[2\overline{11}0]$, $[01\overline{1}0]$ and  $[0001]$ directions are always chosen to be parallel to the $x$, $y$ and $z$ axes.  No periodic boundary conditions were utilized in any direction, which implies that a truly finite-sized NW geometry subject to surface effects was considered in the present work, and that the NW sizes listed in Table \ref{tab:table1} are the actual sizes used for the MD simulations.

\begin{table}
\caption{\label{tab:table1} NW dimensions for all three orientations considered in Fig. \ref{ucori}, where $N_{x}$, $N_{y}$ and $N_{z}$ represent the number of unit cells in each direction.  The aspect ratio for each NW is chosen to be 4:1.}
\begin{center}
\begin{tabular}{|c|c|c|c|c|c|c|c|c||}
\hline
   & \multirow{2}{0.6in}{axial \\ orientation} &  &  &  &  &   & \\ 
   case & &  $N_{x} $ & $N_{y} $  & $N_{z} $   & $l_{x} $  & $l_{y} $    & $l_{z} $   \\
\hline
A1 &  \multirow{3}{0.5in}{$[2\overline{1}\overline{1}0]$ }	 &  30 & 8  & 4 & 9.81 & 2.27 &    2.06  \\
A2 & 										& 42  & 10  & 6 & 13.74 &    2.83 &    3.08   \\
A3 &			 							& 56  & 14  & 8  &  18.32 &    3.97 &     4.11  \\
\hline
B1 & \multirow{3}{0.5in}{$[01\overline{1}0]$} 		& 6 & 32 & 4 &  1.96 &    9.06 &  2.06  \\
B2 & 										 &  10 & 46 & 6 &  3.27 &  13.03  &  3.08  \\
B3 & 								 	& 12 & 60 & 8  & 3.93 & 17.00 &    4.11    \\
\hline
C1 & \multirow{3}{0.5in}{$[0001]$}	 &  6 & 6 & 20 & 1.96 & 1.70 &    10.28   \\
C2 & 							 &  10 & 12 & 30 & 3.27 &   3.40 &  15.42 \\
C3 &  						 &  12 & 14 & 36 & 3.93 &    3.97 &  18.50  \\
\hline
\end{tabular}
\end{center}
\label{tab:size}
\end{table}%

\begin{figure}
\begin{center}
\includegraphics[scale=0.5]{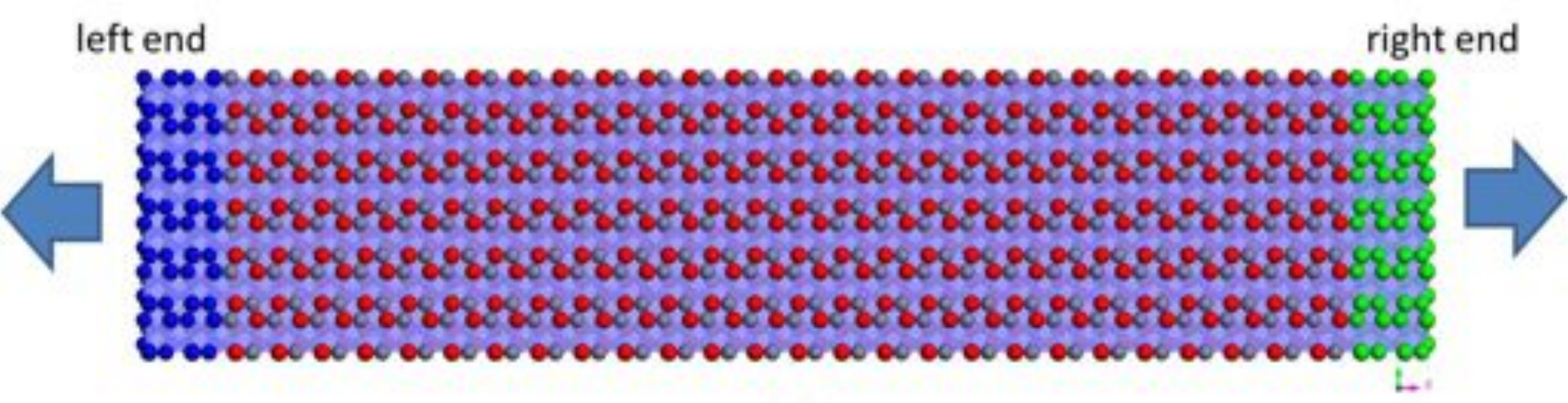}
\caption{Illustration of fixed (left end (blue) and right end (green)) and free (red and gray) atoms for axial loading.  The NW shown has a $[0001]$ axial direction.}
\label{axialbending}
\end{center}
\end{figure}

We performed MD simulations of tensile axial deformation.  For the tensile loading, both ends of the NW were first allowed to relax dynamically to a new equilibrium length in response to surface stresses by using a Berendsen thermostat~\cite{Berendsen:1984us} for up to 400 ps depending on the NW cross sectional size.  After the new equilibrium length was found, two unit cells at each end of the NW, as illustrated in Fig. \ref{axialbending}, were held fixed while the NW was equilibrated using a Nose-Hoover thermostat~\cite{hooverPRA1985} for 20 ps.  After these two initial equilibrium steps, the ends of the NW were displaced axially at strain increments of 0.25\% and held fixed while the NW was relaxed for 20 ps.  After each strain increment, both ends of the NW were held fixed while the remainder of the NW was dynamically equilibrated for 100 ps using the Nose-Hoover thermostat at a temperature of 300K.  The loading was increased until an axial strain of 20\% in tension was reached.  

\begin{figure}
\begin{center}
\includegraphics[width=4.3in]{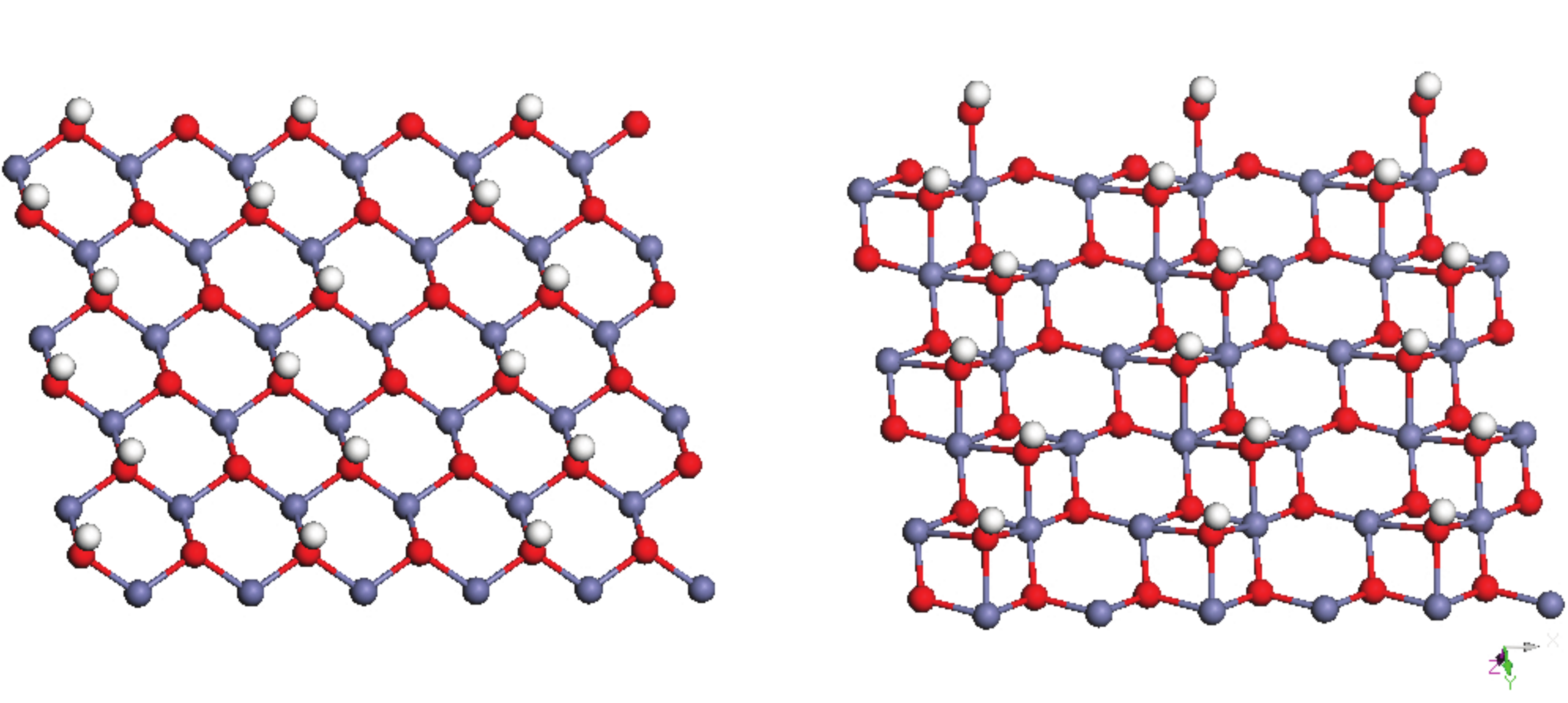}
\caption{Illustration of surface passivation via (left) adsorption of an H atom on O terminated $(000\bar{1})$, and (right) OH adsorbed on the Zn terminated $(0001)$ surface leading to a $2\times1$ pattern for both surfaces.  (red: O, grey: Zn, white: H)}
\label{fig:surpas}
\end{center}
\end{figure}

For the surface treatments to avoid the electrostatic divergence, we note that the surface atoms on the (0001) polar surfaces of ZnO have three nearest neighbors instead of four as does a bulk Zn or O atom.  For the SP treatment, we added an H atom to the O-terminated surface, and added an OH molecule to the Zn terminated surface in order to saturate dangling bonds, which, as illustrated in Fig. \ref{fig:surpas}, results in a 2$\times$1 passivation on both the O and Zn-terminated polar (0001) surfaces.  The 2$\times$1 passivation is used in our work for multiple reasons.  First, as previously mentioned it has been observed experimentally~\cite{Lauritsen:2011db} and has been used in previous DFT calculations~\cite{Dag:2011hg,Wander:2001wh}.  Furthermore, it is also the smallest passivation pattern that we can use on the NW polar surfaces due to the relatively small cross-sectional sized NWs we consider in this work.  The potential parameters for both the Zn-O interactions as modeled using the potential of~\cite{binksSSC1994}, as well as the parameters for the H-O interactions needed for the surface passivation as taken from~\cite{deLeeuw:1998wk} are listed in Table \ref{tab:para}.  This passivation is also realistic as it is common for the environment to contain some water or humidity; the effects of water on the elastic properties of ZnO have also recently been investigated~\cite{Yang:2011gl}.  We note the likelihood that the piezoelectric properties of ZnO will depend on the specific passivation that is utilized in computation, or that occurs experimentally.

\begin{table}
\caption{\label{tab:para}Buckingham parameters for the Zn-O interactions from~\cite{binksSSC1994}, and also for the H-O interactions taken from~\cite{deLeeuw:1998wk} for the surface passivation.}
\begin{tabular}{cccc}
\hline
 Species & A (eV) & $\rho$ (\AA) & C (eV\AA$^6$)  \\
\hline
 	O$^{2-}$-O$^{2-}$ 	& 9547.96 	& 0.21916 	& 32.0  \\
       	 Zn$^{2+}$-O$^{2-}$ 	& 529.70 		& 0.3581 		& 0.0   \\
	 Zn$^{2+}$-Zn$^{2+}$ 	& 0.0		 	& 0.0 		& 0.0   \\
\hline
	 H$^{+}$-O$^{2-}$ 	& 396.27  	& 0.25    	& 0.0  \\
\hline
\end{tabular}
\end{table}

For the surface treatment using CC, the methodology is much more straightforward.  Because each Zn and O atom on a polar (0001) surface has 75\% of the neighbors of the corresponding bulk atom (i.e. 3 instead of 4), the charge of the top layer of Zn and O atoms is reduced to 75\% of the formal charge from $\pm2e$ to $\pm1.5e$~\cite{nogueraJPCM2000,daiJAP2011}.  The CC surface treatment can also be physically justified as enforcing partial covalence of the surface atoms as compared to bulk atoms.  We also note that the SP and CC surface treatments can be used for other polar crystals~\cite{AvraamPRB2011,Stengel:2011gj}.

The key value of interest we will report is the change in polarization for the NW as a function of the applied mechanical deformation.  This parameter is key for design of nanogenerators as a larger polarization is directly related to a higher output voltage~\cite{Shao:2010jw,gaoNL2007,Sun:2010cm}, and thus more electrical energy generation~\cite{Kamel:2010bf,Yan:2011et,Sun:2010cm}.  For the axial loading, we accomplish this by calculating, for each state of strain, the polarization of each unit cell, then summing over the entire NW to calculate the total NW polarization, where the unit cell is defined by the group of four atoms in the green box in Fig. \ref{ucori}(a), i.e.:
\begin{equation}
\label{eqn:unicellP}
P_{cell}=\sum^{4}_{i=1 }\frac{z_{i}q_{i}}{\Omega_{cell}}, P_{3}=\sum^{N}_{j=1} P_{cell}/N,
\end{equation}
where $P_{cell}$ is the polarization for a single unit cell, $P_{3}$ is the polarization of the NW in the polar $[0001]$ direction, $N$ is the total number of unit cells in the system, $q$ is the charge on each atom and $z$ is the coordinate of each atom.  We note that the effective piezoelectric constants are calculated using the invariant definition of~\cite{vanderbiltJPCS2000}:
\begin{equation}
e^{\text{eff}}_{31}=\frac{1}{2}\frac{d P_3}{d \varepsilon_{1}}+P_3, e^{\text{eff}}_{32}=\frac{1}{2}\frac{dP_3}{d \varepsilon_{2}}+P_3, e^{\text{eff}}_{33}=\frac{d P_3}{d \varepsilon_{3}}.
\end{equation}

\section{Numerical Results}

\subsection{Mechanical Properties}

\begin{figure}
\begin{center}
\includegraphics[width=3.9in]{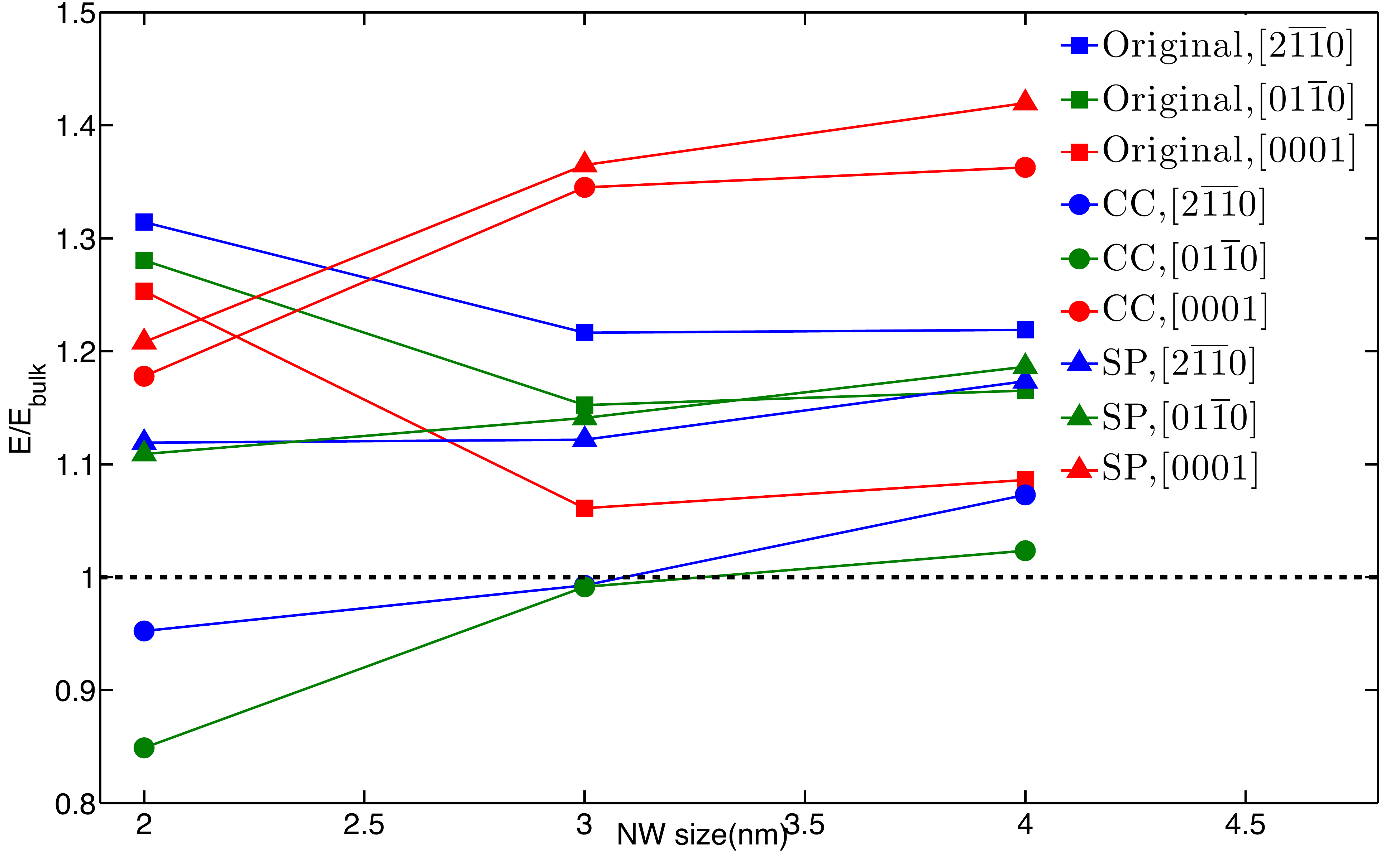}
\caption{Bulk-normalized Young's modulus for the three NW orientations and geometries summarized in Table \ref{tab:table1} for CC, SP, and original (untreated) surface treatments.}
\label{fig:ECCSP}
\end{center}
\end{figure}

We first discuss the effect that the different surface treatments (CC vs. SP) have on the elastic properties of ZnO NWs, where the results are summarized in Fig. \ref{fig:ECCSP}.  The Young's modulus for each orientation was calculated by normalizing by the bulk Young's modulus for each orientation, which were taken to be 156.2 GPa for the $[2\overline{1}\overline{1}0]$ and $[01\overline{1}0]$ orientations and 119.7 GPa for the $[0001]$ orientation~\cite{Lee:2003js}.  The results are consistent for all three NW orientations considered:  the Young's modulus is highest when no surface treatment is performed (original), followed by SP followed by a substantial reduction for the CC case.  Furthermore, the modulus is observed to increase with decreasing size for all three orientations for the original case, which is consistent with previous MD simulation reports on ZnO NW elastic properties~\cite{kulkarniNANO2005}, whereas a size-dependent decrease in Young's modulus is observed for both the CC and SP cases.  The Young's modulus is lowest for the CC case because of the reduction in formal charge for the surface atoms, which leads to reduced interaction energies, forces and thus stiffness for the surface atoms, and for the $[0001]$ and $[01\overline{1}0]$ orientations leads to a Young's modulus that is smaller than the bulk value for the smallest NW sizes we considered.  The modulus for the SP case is similar to the no treatment case, but is typically slightly smaller and is found to be larger than the bulk value for all sizes considered.  

\begin{figure}
\centering
\subfigure[]{
\includegraphics[width=4.0in]{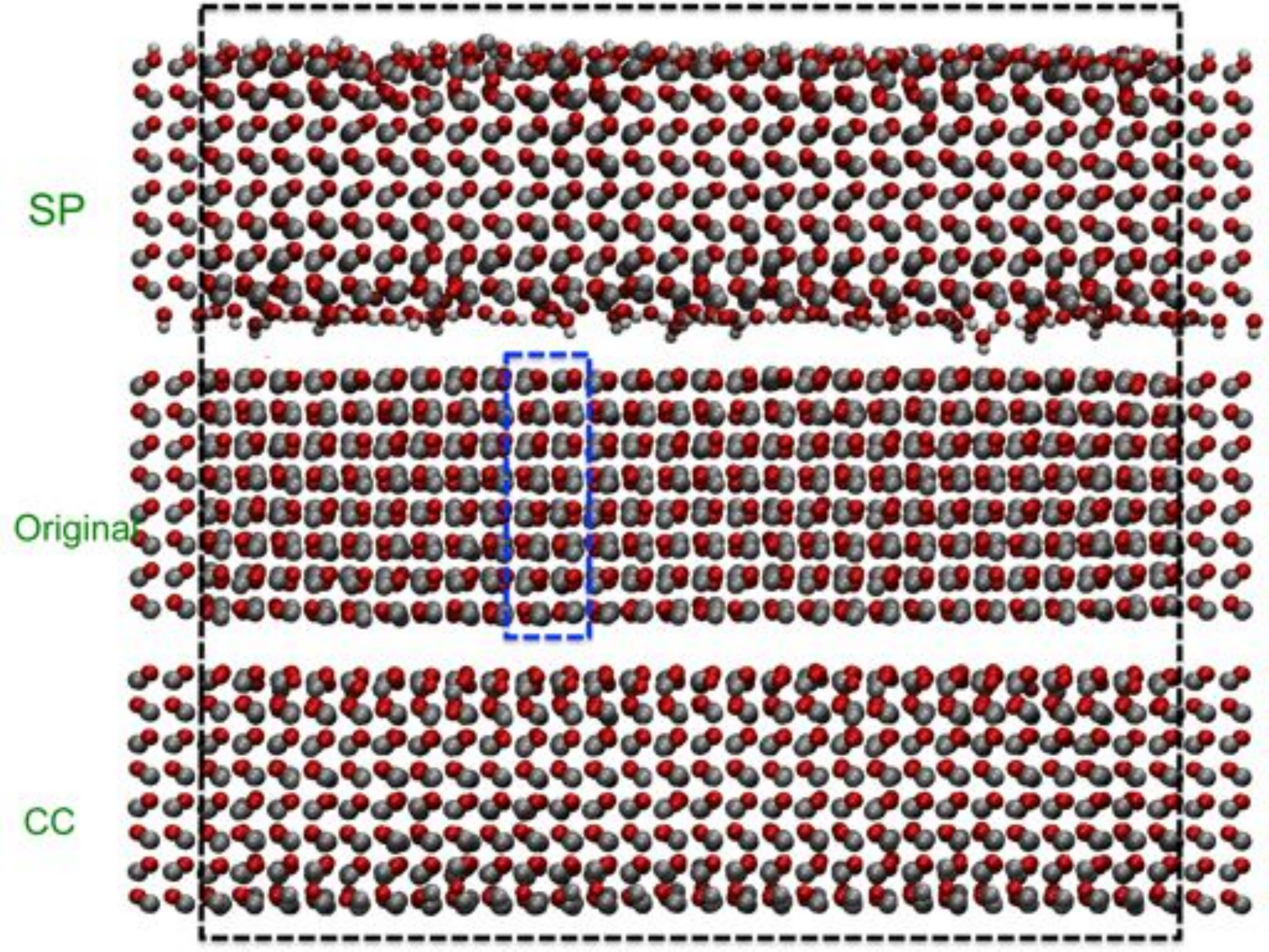}
\label{fig:com3yconf}
}

\centering
\subfigure[]{
\includegraphics[width=4.0in]{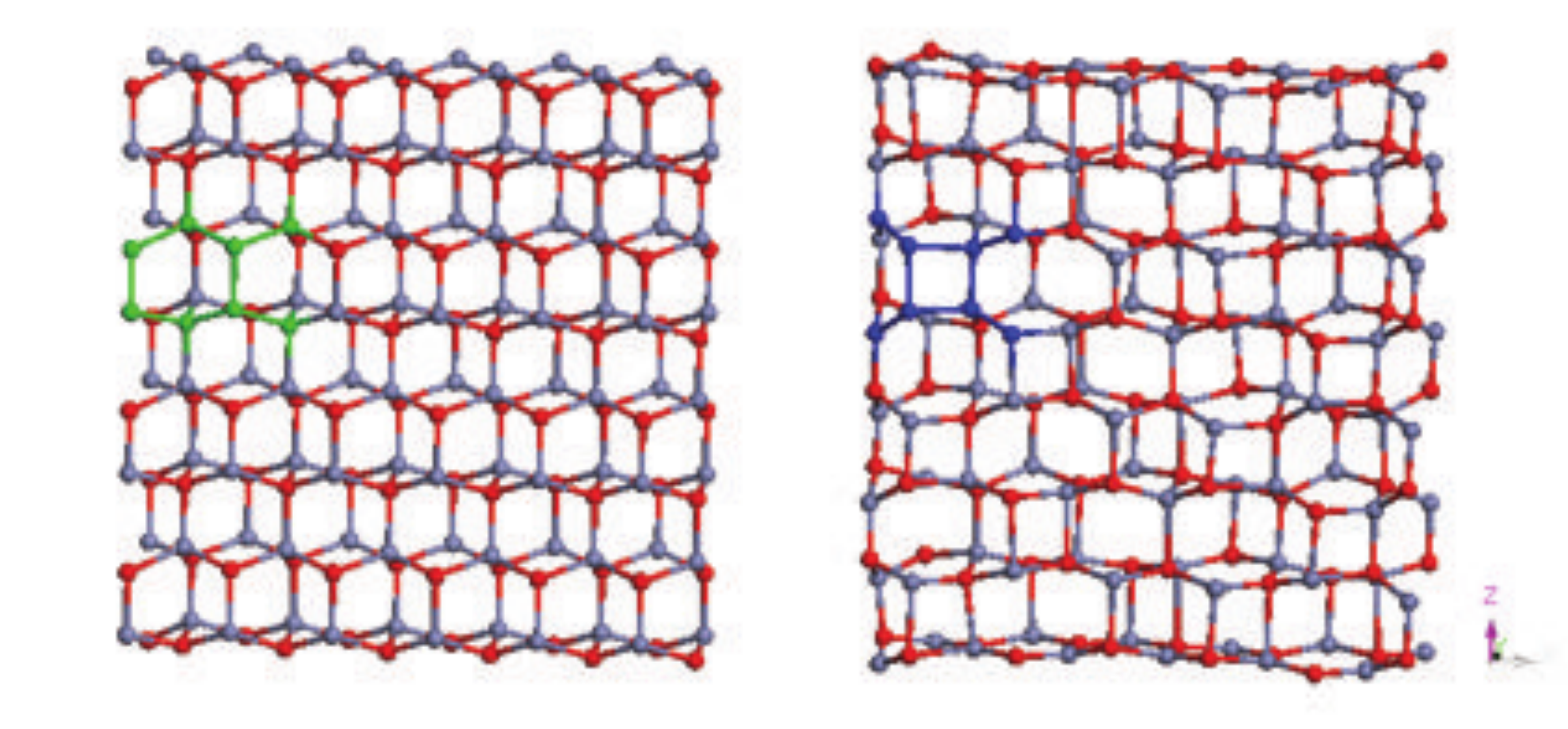}
\label{fig:com3yconftran}
}
\caption{(a): Snapshot of the atomic configuration at zero tensile strain for SP, original and CC surface types for the $[01\overline{1}0]$ orientation showing that the SP and CC NWs keep the original WZ lattice structure, while the original NW has transformed to a d-BCT phase.  (b) Comparison of the (left) WZ lattice structure to the (right) d-BCT lattice structure (taken from the blue rectangular box in (a)), where two unit cells are chosen and highlighted in blue for comparison to show the typical rectangular structure of the d-BCT phase.}
\label{fig:0110}
\end{figure}

\begin{figure}
\centering
\subfigure[]{
\includegraphics[width=4.0in]{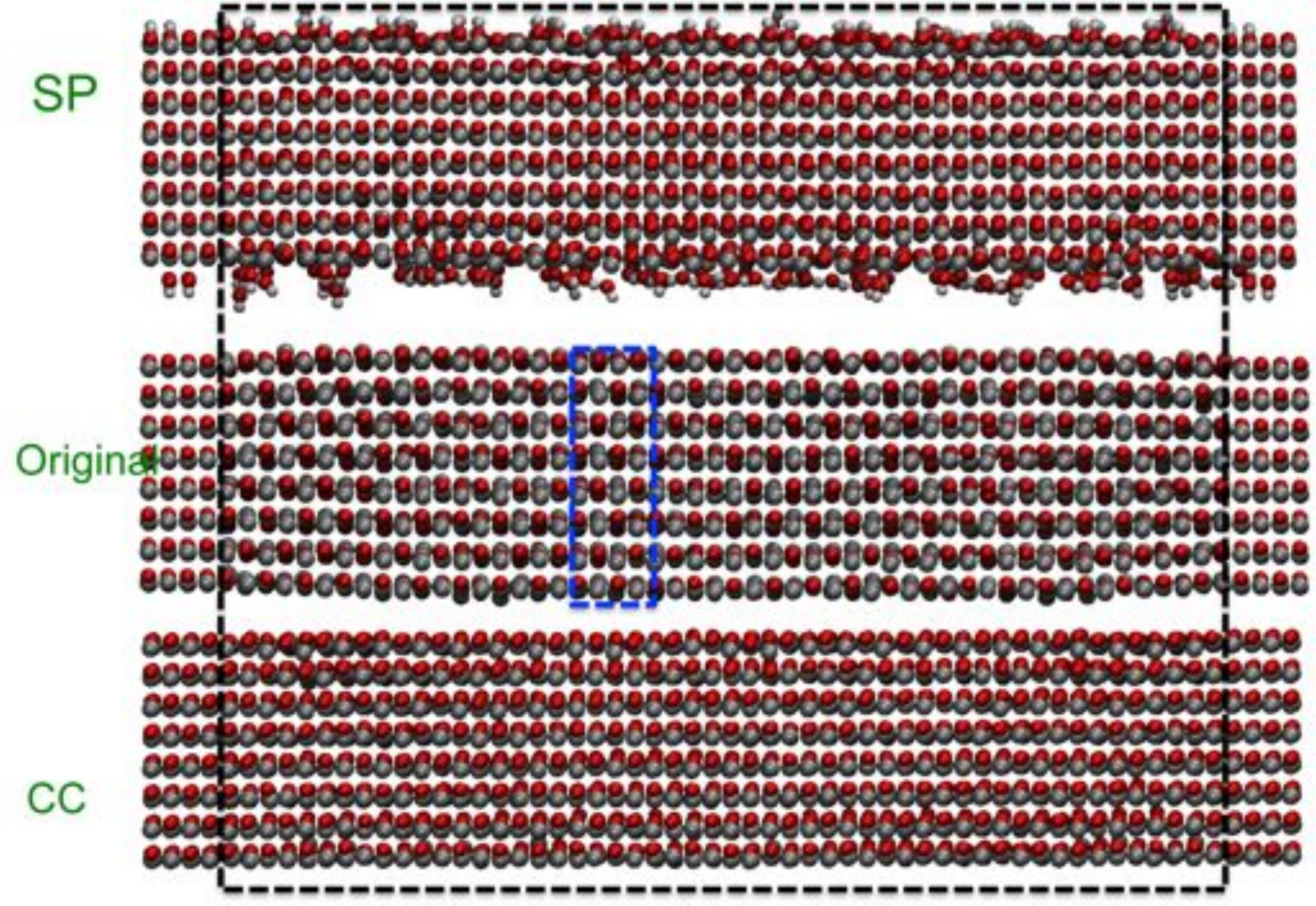}
\label{fig:com3xconf}
}

\centering
\subfigure[]{
\includegraphics[width=4.0in]{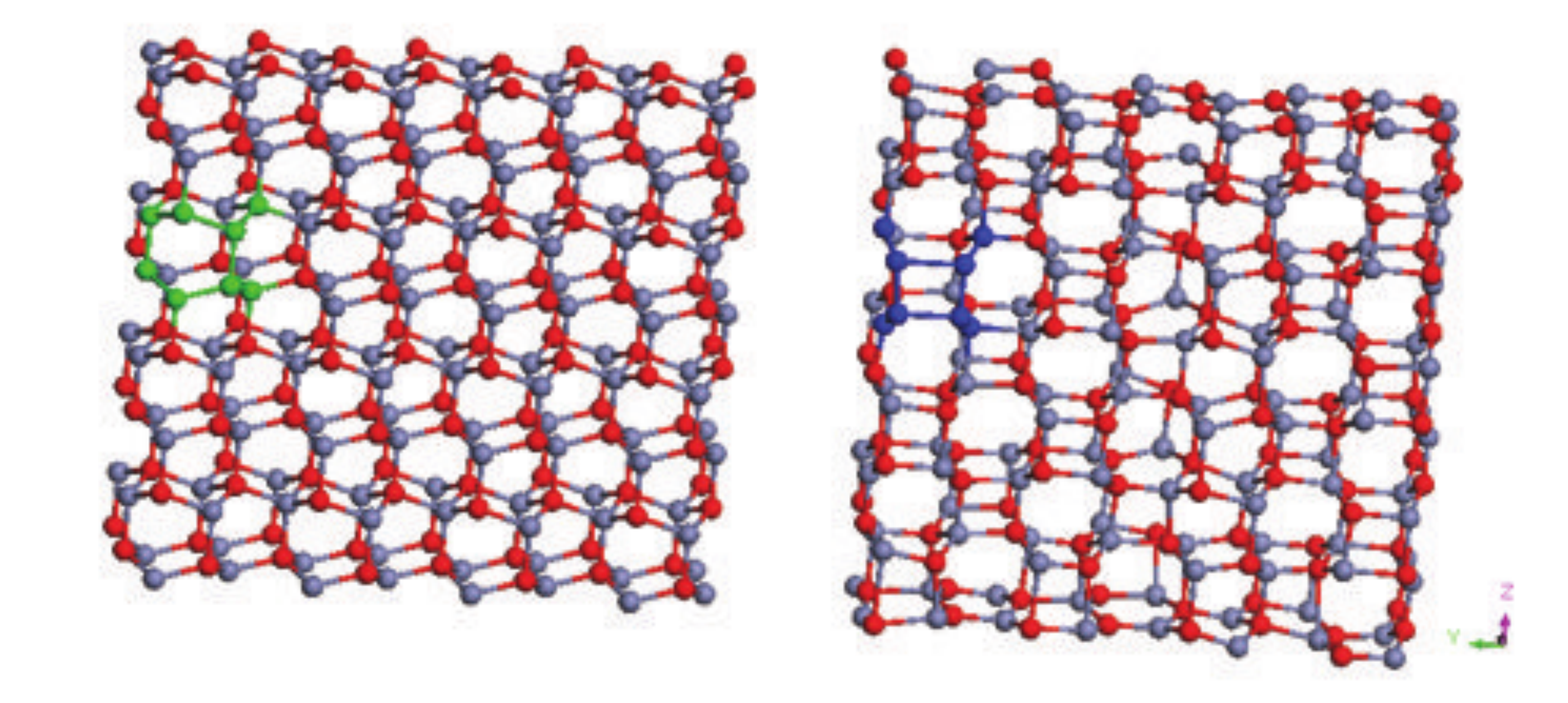}
\label{fig:com3xconftran}
}
\caption{(a): Snapshot of the atomic configuration at zero tensile strain for SP, original and CC surface types for the $[2\overline{11}0]$ orientation showing that the SP and CC NWs keep the original WZ lattice structure, while the original NW has transformed to a d-BCT phase.  (b) Comparison of the (left) WZ lattice structure to the (right) d-BCT lattice structure (taken from the blue rectangular box in (a)), where two unit cells are chosen and highlighted in blue for comparison to show the typical rectangular structure of the d-BCT phase.}
\label{fig:2110}
\end{figure}

While the Young's modulus trends and values in Fig. \ref{fig:ECCSP} may seem surprising, the atomistic origin of these trends can be observed as shown in Figs. \ref{fig:0110} and \ref{fig:2110}, where the atomic structure before any axial loading is applied is shown for both the $[01\overline{1}0]$ and $[2\overline{11}0]$ orientations.  Specifically, it is shown that for both the $[01\overline{1}0]$ and $[2\overline{11}0]$ orientations, if no surface treatment is performed, as has been the case in previous calculations~\cite{Morgan:2007gb, Morgan:2009cp, Morgan:2010fl}, the WZ lattice structure is unstable and transforms to a d-BCT structure.  In fact, this transformation occurs during the initial relaxation phase of the simulation for all NW sizes we have considered, and therefore the Young's moduli for the $[01\overline{1}0]$ and $[2\overline{11}0]$  orientations in Fig. \ref{fig:ECCSP} correspond to that of the d-BCT, and not WZ structure.  We note that regardless of the surface treatment, no initial transformation occurs for the $[0001]$ NWs because the polar direction is along the axial direction for this case.

In contrast, the SP and CC cases for both the $[01\overline{1}0]$ and $[2\overline{11}0]$ orientations result in a stable WZ structure for the NW sizes we have considered, which demonstrates if the polarization divergence due to the polar $(0001)$ surfaces is not treated, the WZ lattice structure is not stable and transforms to the d-BCT structure.  The transformation to the d-BCT phase will have significant ramifications on the piezoelectric constants, as we will discuss shortly.  Before continuing to that discussion, we note that after yield, the NWs transform into a non-piezoelectric structure~\cite{Kulkarni:2008kp, Sarasamak:2008vh, Wang:2008ki, Wang:2007gd, kulkarniNANO2005}.  Specifically, the $[01\overline{1}0]$ NW transforms to a hexagonal~\cite{kulkarniPRL2006} phase, while the $[0001]$ orientation transforms to the d-BCT phase~\cite{Agrawal:2010go}.  The transformation to a non-piezoelectric phase can also be observed by the post-yield behavior in the polarization vs. strain curves in Figs. \ref{fig:ps2110}, \ref{fig:ps0110} and \ref{fig:ps0001}, which we discuss next in further detail.

\subsection{Piezoelectric Constants}

\begin{figure}
\begin{center}
\includegraphics[width=4.0in]{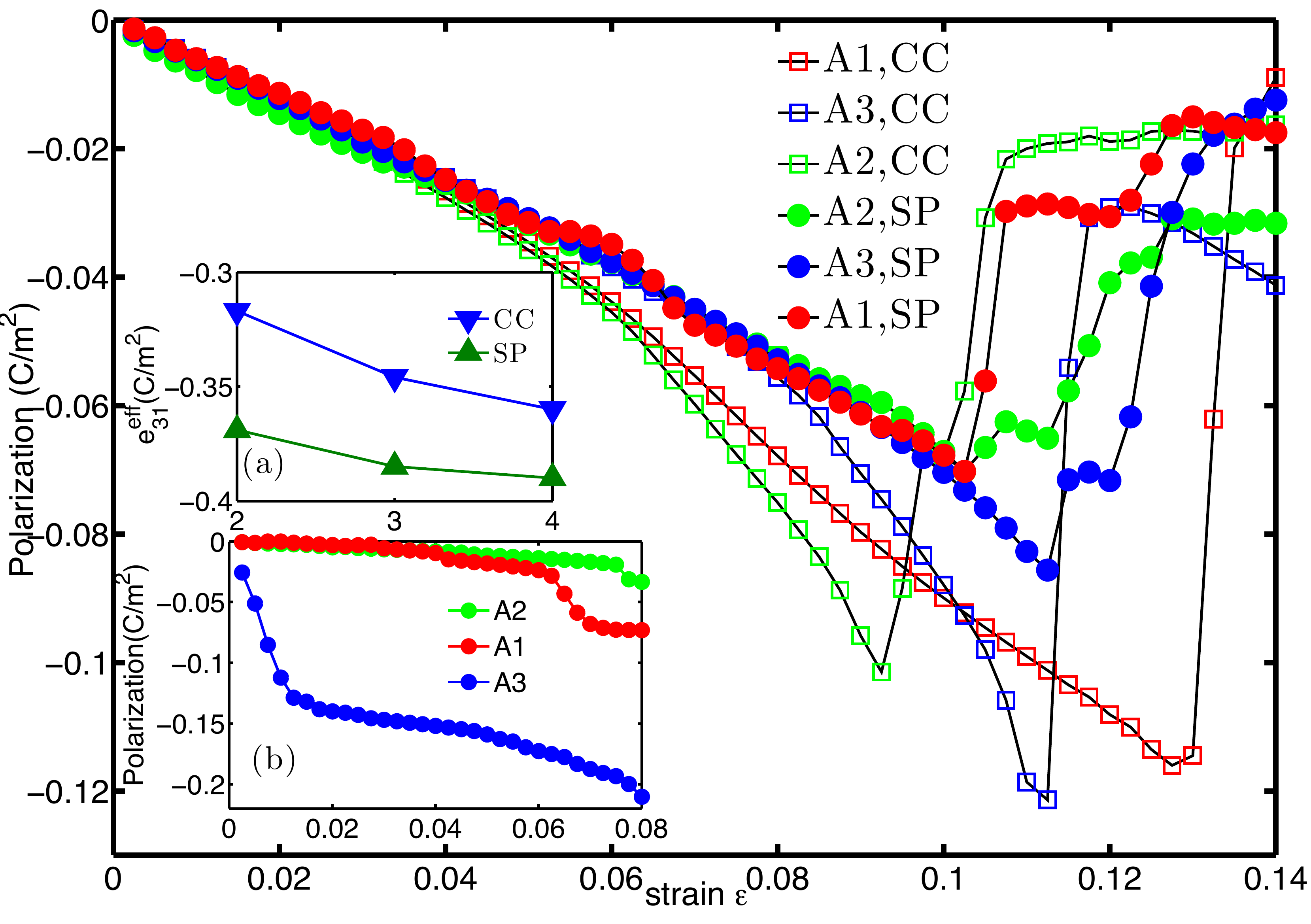}
\caption{Polarization vs. strain for axial loading along the $[2\overline{1}\overline{1}0]$ direction for different NW sizes. Inset (a) Size-dependent effective piezoelectric coefficient $e_{31}^\text{{eff}}$; (b) Polarization vs. strain for original (untreated surface) NW.}
\label{fig:ps2110}
\end{center}
\end{figure}

\begin{figure}
\begin{center}
\includegraphics[width=4.0in]{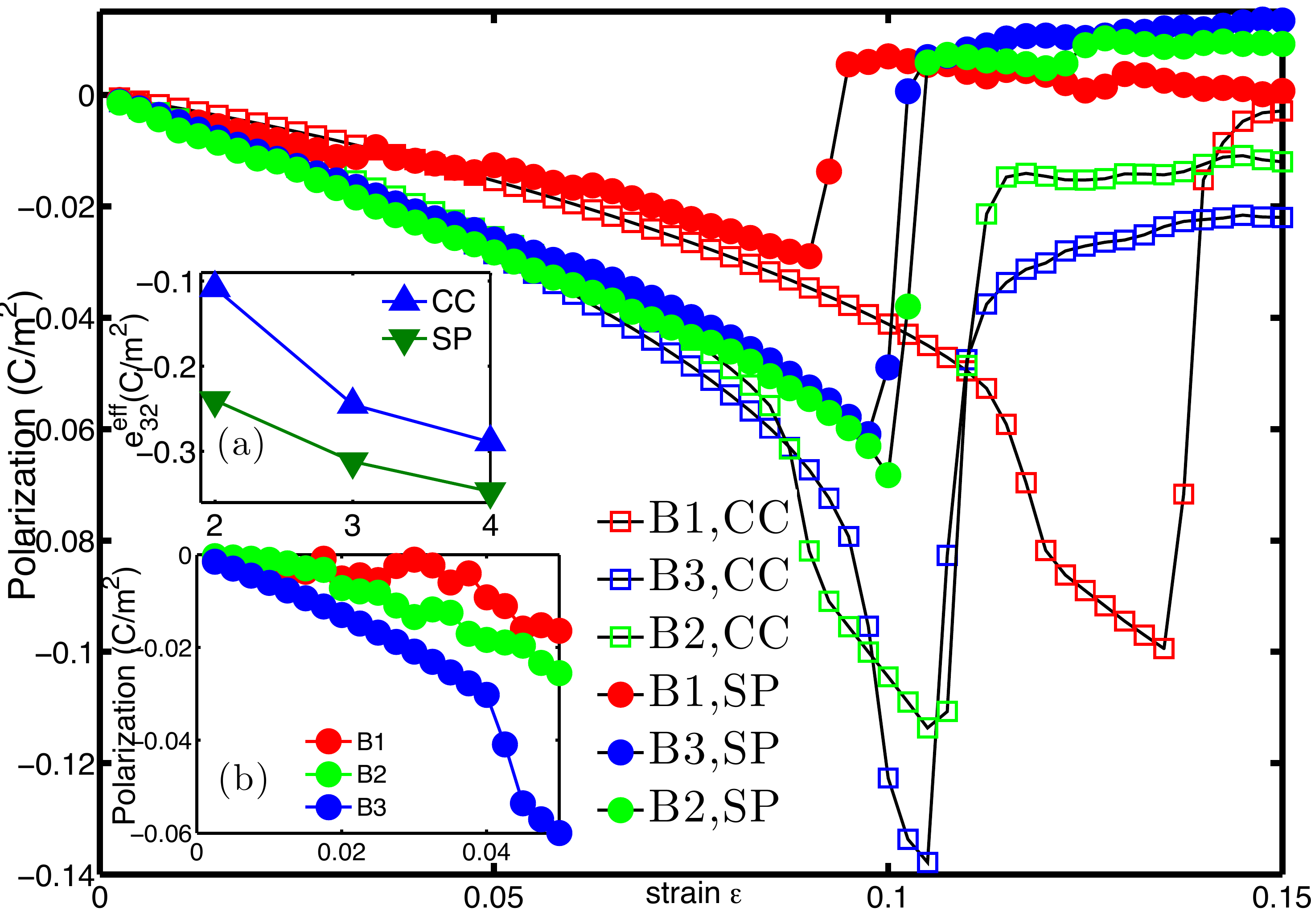}
\caption{Polarization vs. strain for axial loading along the $[01\overline{1}0]$ direction for different NW sizes. Inset (a) Size-dependent effective piezoelectric coefficient $e_{32}^\text{{eff}}$; (b) Polarization vs. strain for original (untreated surface) NW.}
\label{fig:ps0110}
\end{center}
\end{figure}

\begin{figure}
\begin{center}
\includegraphics[width=4.0in]{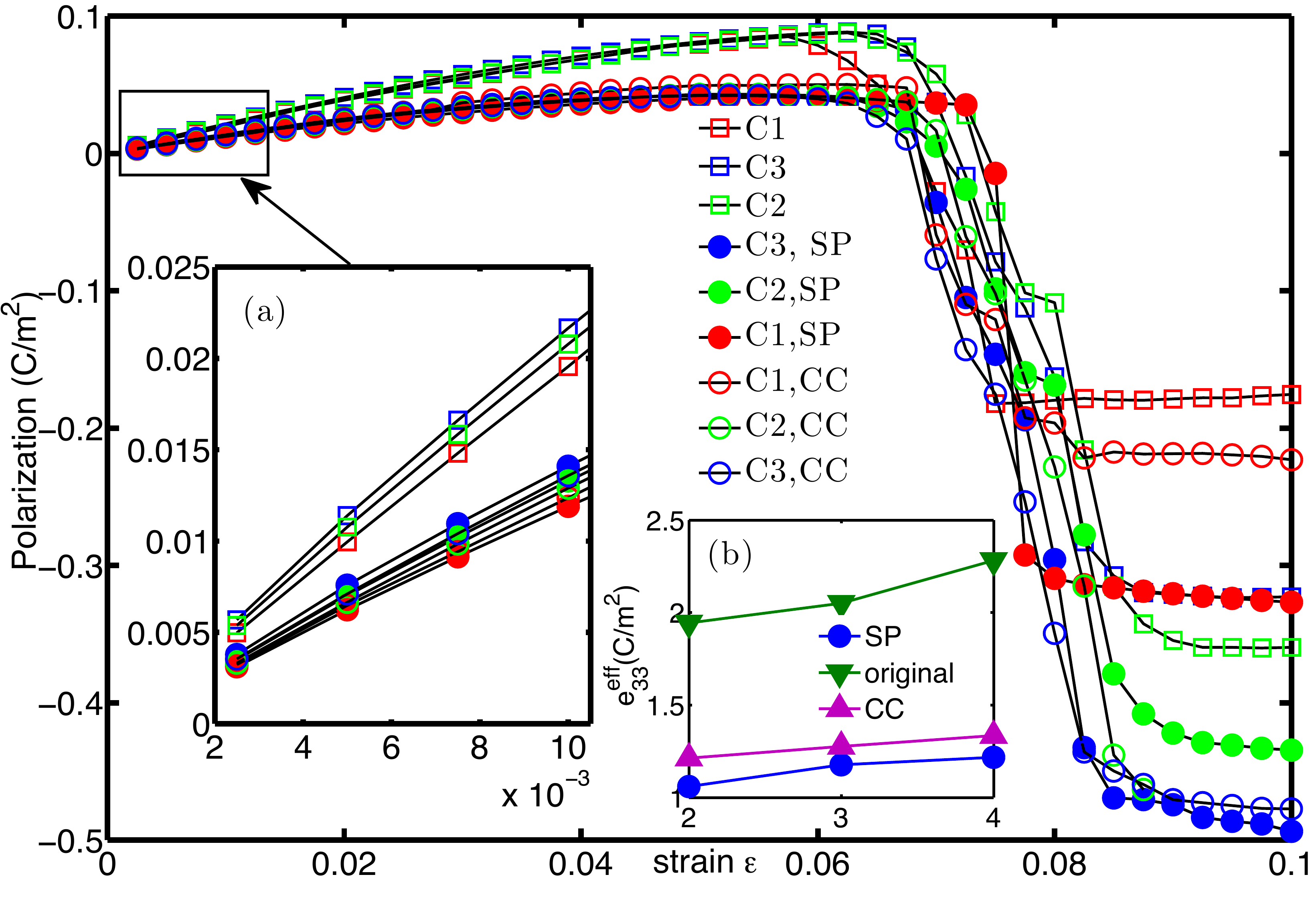}
\caption{Polarization vs strain for axial loading along the $[0001]$ direction for different NW sizes.  Inset (a) Zoom in to the small strain regime; (b) Size-dependent effective piezoelectric coefficient $e_{33}^\text{{eff}}$.}
\label{fig:ps0001}
\end{center}
\end{figure}

The polarization vs. strain for all three NW orientations is shown in Figs. \ref{fig:ps2110}, \ref{fig:ps0110} and \ref{fig:ps0001}.  The first issue to point out is that, if CC or SP is utilized for the $[01\overline{1}0]$ and $[2\overline{11}0]$ orientations in Figs. \ref{fig:ps2110} and \ref{fig:ps0110}, the polarization is linearly dependent on strain until yield, which occurs around 10\% tensile strain for both orientations.  However, as seen in Figs. \ref{fig:ps2110}(b) and \ref{fig:ps0110}(b), if no surface treatment is utilized, the slope of the polarization vs. strain curve varies significantly even at very small amounts of applied tensile strain.  Furthermore, due to the transformation from the WZ to non-piezoelectric d-BCT structure, as shown in Figs. \ref{fig:0110}(b) and \ref{fig:2110}(b), the polarization vs. strain curves are quite noisy and do not converge with decreasing size, which is a direct result of the non-piezoelectric d-BCT structure.  We note that for the $[2\overline{11}0]$ orientation, the relevant effective piezoelectric constant is $e_{31}^{\text{{eff}}}$, while for the $[01\overline{1}0]$ orientation, the relevant effective piezoelectric constant is $e_{32}^{\text{{eff}}}$.

The $[0001]$ orientation exhibits a different polarization vs. strain response than the $[01\overline{1}0]$ and $[2\overline{11}0]$ orientations because, as previously discussed, the WZ phase is stable without any surface treatment.  Therefore, as shown in Fig. \ref{fig:ps0001}, the slope of the polarization vs. strain curve is linear even for the original surface case.  However, it is clear that the slope for the original surface is significantly higher than the slopes for the CC and SP cases.  A simple argument as to why the original, untreated surface leads (incorrectly) to a larger piezoelectric constant can be given as follows: the dipole moment for the original NW with untreated surfaces can be estimated as $M_{\text{Original}}=2*2e*d_{1}N$, where $d_{1}$ and $d_{2}$ are the layer distances that are related by $d_{1}=d_{2}\frac{u}{0.5-u}$, $N$ is number of unit cells in the polar $[0001]$ direction. The dipole moment of the CC NWs is $M_{\text{CC}}=-2*2e*d_{2}(N-1)+2*1.5e(Nd_{1}+(N-1)d_{2})$, so $\frac{M_{\text{Original}}}{M_{\text{CC}}}\cong \frac{8u}{8u-1}=1.475$.   This explains why in Table \ref{tab:eijeff} the original $e^{\text{eff}}_{33}$ is larger than $e^{\text{eff}}_{33}$ from CC and SP, and demonstrates the necessity of CC and SP to obtain the more reasonable effective piezoelectric constants.

A summary of the effective piezoelectric constants for all orientations and sizes is given in Table \ref{tab:eijeff}, where a comparison of the bulk piezoelectric constants as calculated for the~\cite{binksSSC1994} potential are given for reference.  As can be seen, there is a decrease in effective piezoelectric constant with decreasing size if the CC and SP are utilized.  Our previous MD and DFT study~\cite{daiJAP2011} also found that the effective piezoelectric coefficients of ZnO thin film decrease as the film thickness decreases.  The decrease in $e_{33}^\text{{eff}}$ is less dramatic, with the reduction reaching 16.6\% for the smallest NW sizes considered.  In contrast, the reduction in $e_{31}^\text{{eff}}$ and $e_{32}^\text{{eff}}$ is more dramatic, reaching 80.0\% for the smallest NW sizes considered for $e_{32}^\text{{eff}}$.  As found before for the Young's modulus, the reduction in the piezoelectric constants is generally larger if CC is utilized as compared to SP.  

\begin{table}
\caption{Summary of effective piezoelectric constants (units of C/m$^{2}$) for the different NW sizes and orientations from Table \ref{tab:table1}.  Comparison with the bulk piezoelectric constants from~\cite{daiNANO2010} provided for reference. Labeling for NWs is the same as in Table \ref{tab:table1} for consistency. }
\begin{center}
\begin{tabular}{|c|c|c|c|c|c|c|c|c|c|}
\hline
& size &  $e_{33}^{\text{eff}}$ & size  & $e_{31}^{\text{eff}}$  &  size & $e_{32}^{\text{eff}}$     & & size & $e_{33}^{\text{eff}}$  \\	
\hline	
\multirow{3}{*}{CC} 	& C1	 		& 1.213 		& A1		& -0.317 & B1	& -0.108		& \multirow{3}{*}{Original}  & C1 & 1.94 	\\
  				&  C2		& 1.276	 	& A2 	& -0.346 &  B2	& -0.245 		& 					& C2	 & 2.05 	\\ 
 				& C3 		& 1.335 		& A3 	& -0.360 & B3	& -0.289		&  				& C3		& 2.28 	\\
\hline	
\multirow{3}{*}{SP}  	&  C1	 	& 	1.059	& A1 	& -0.369		& B1	& -0.240		 &	  	&				& 	\\
 				& C2			& 	1.177	 & A2	& -0.385		& B2	& -0.312			&   	&			& 	\\ 
  				& C3			& 	1.217	 & A3 	& -0.390		& B3 	&  -0.347		&   	&		& 	\\
\hline
  Bulk  & 		   &      1.27         &    & -0.54           &     &      -0.54            	&  		&				&  \\
\hline
\end{tabular}
\end{center}
\label{tab:eijeff}
\end{table}%

\begin{figure}
\centering
\subfigure[]{
\includegraphics[width=4.0in]{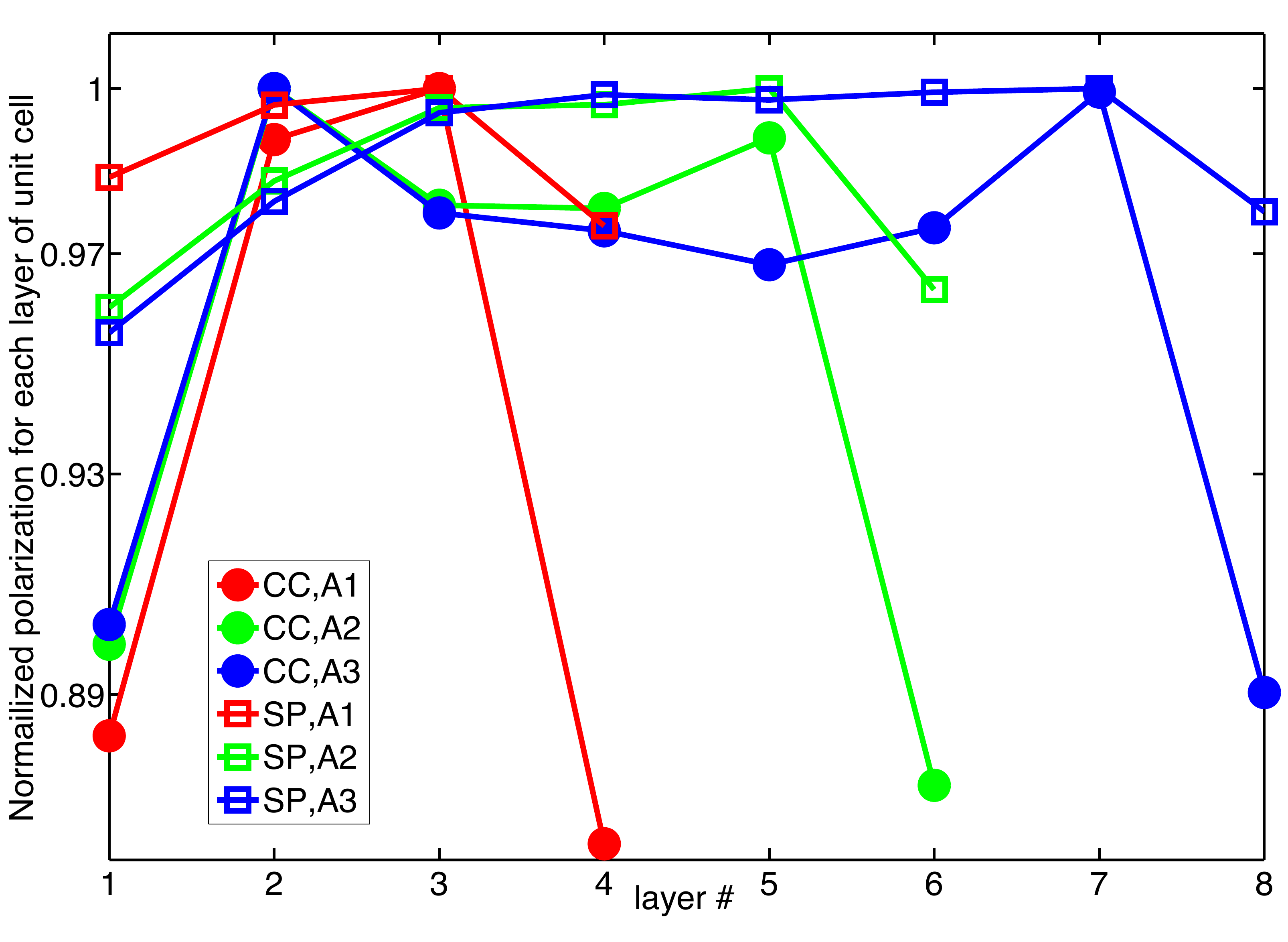}
\label{fig:ucpX}
}
\centering
\subfigure[]{
\includegraphics[width=4.0in]{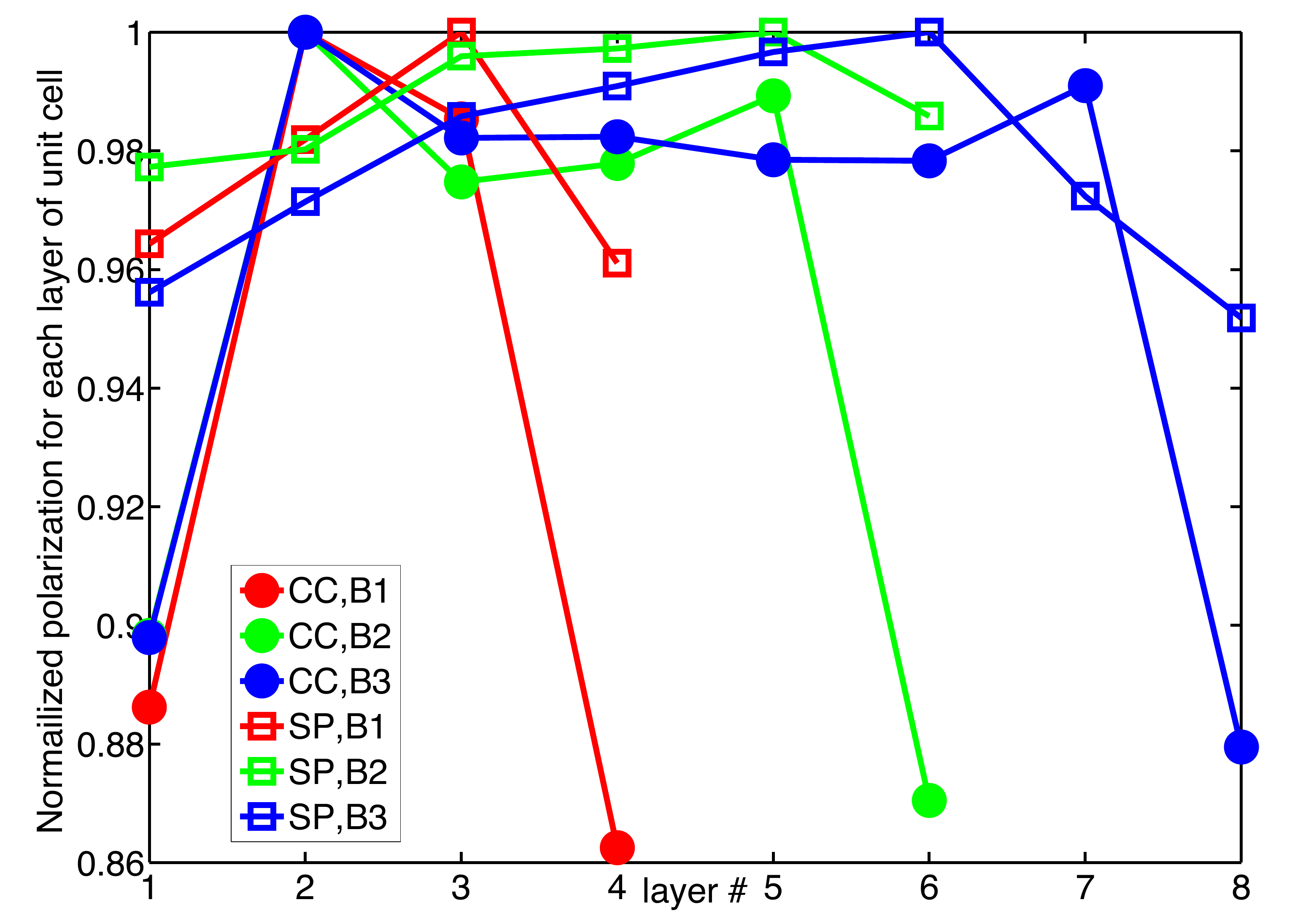} 
\label{fig:ucpY}
}
\caption{Variation in unit cell polarization through the NW [0001] direction for both the \subref{fig:ucpX} $[2\overline{11}0]$ and \subref{fig:ucpY} $[01\overline{1}0]$ axial orientations, which demonstrates the reduction in surface unit cell polarization as compared to the bulk.  The polarization of each unit cell is normalized by the polarization of a bulk unit cell. }
\label{fig:ucthickness}
\end{figure}

We now address the mechanism underlying the smaller piezoelectric constants that we have found for the CC and SP surface treatments.  The approach we take, similar to previous works~\cite{Behera:2008ty,zhangNANO2009,zhangNANO2010}, is to analyze the polarization on a unit-cell basis through the polar [0001] direction of the NWs.  For the $[2\overline{11}0]$ and $[01\overline{1}0]$ orientated NWs, the variation in unit cell polarization through the NW polar [0001] direction is shown in Fig. \ref{fig:ucthickness}, where the polarization at the surfaces corresponds to the polar (0001) surfaces.  As shown in Fig. \ref{fig:ucthickness}, the unit cell polarization at the surface is reduced by more than 10\% for the CC surface treatment for both NW orientations, while the surface unit cell polarization reduction is smaller, i.e. less than 5\% for all SP-oriented NW sizes.  The polarization reduction is largest at the polar surface for both CC and SP, then converges to the bulk value as the interior of the NW is reached.  

\begin{figure}
\centering
\includegraphics[width=4.0in]{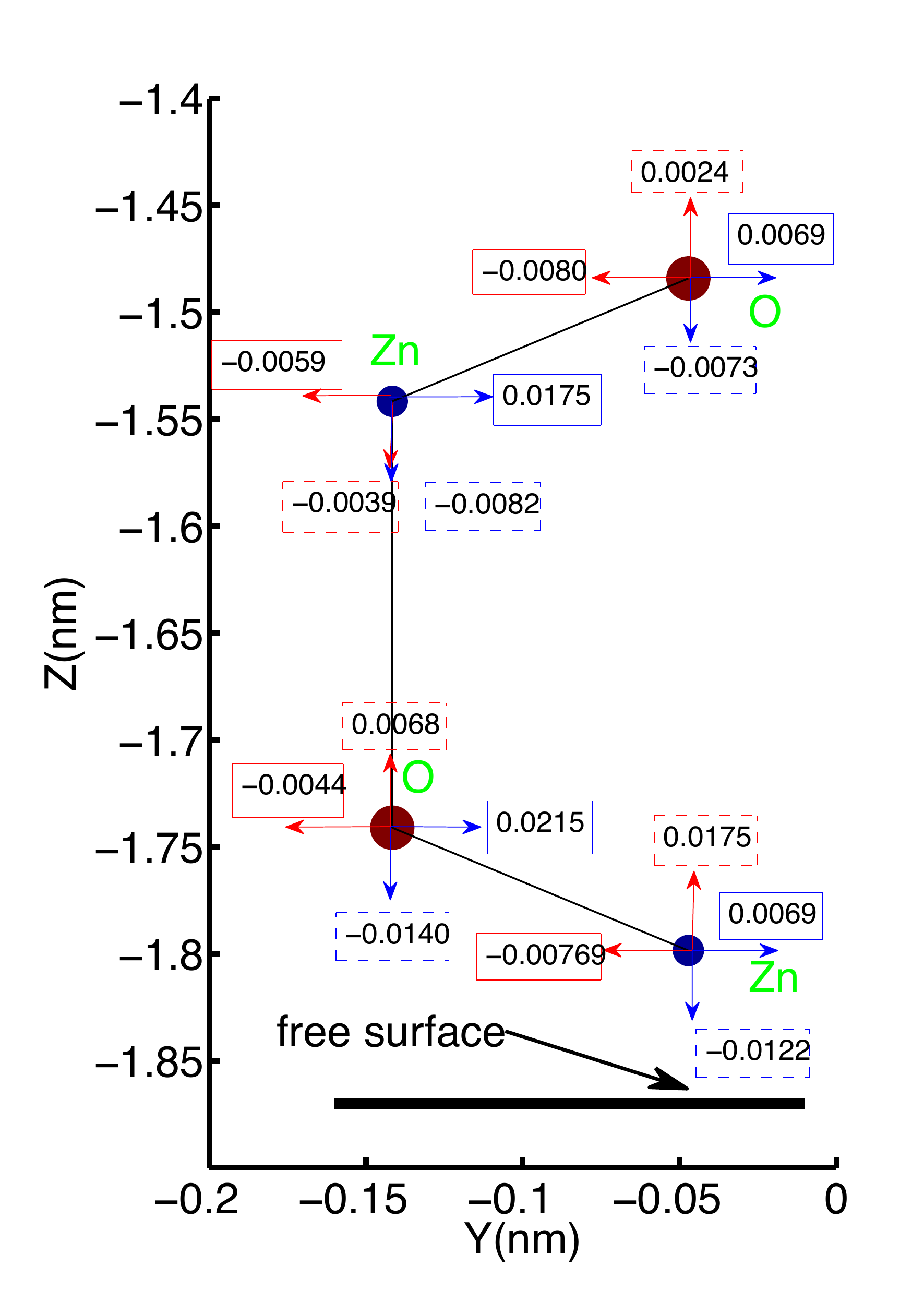}
\caption{Side view of the surface unit cell (four atoms, or two Zn-O dimers) for the $[2\overline{11}0]$ oriented ZnO NW with size A3.  The red and blue arrows show the atomic displacements for CC and SP, respectively, with the values labeled in the corresponding boxes.  The image shows that while the surface atoms contract for CC, but expand for SP in the [0001] direction, the relative distance between the Zn and O atoms that comprise the dimer nearest to the surface becomes smaller for both CC and SP.}
\label{fig:surfatoms}
\end{figure}

The atomistic deformation leading to the reduction in polarization for the surface unit cells as shown in Fig. \ref{fig:ucthickness}, and thus the reduction in effective piezoelectric constant as shown in Table \ref{tab:eijeff} is shown in Fig. \ref{fig:surfatoms}, which shows a snapshot of a surface unit cell for the $[2\overline{11}0]$ oriented ZnO NW with size A3, with atomic displacements resulting from both CC and SP surface treatments.  It is at first glance surprising that both surface treatments lead to decreases in the effective piezoelectric constants with decreasing NW cross sectional size because the surface contracts for CC and expands for SP in the [0001] direction in response to surface stresses as shown in Fig. \ref{fig:surfatoms}.  However, the important parameter for the polarization is not the absolute displacement of atoms near the surface, but instead the relative displacements between the Zn-O atoms that comprise each of the two dimers in the surface unit cell, as can be seen through inspection of Eq. \ref{eqn:unicellP}.  Specifically, the relative distance between the Zn-O dimer closest to the surface in Fig. \ref{fig:surfatoms} is -0.0107 nm and -0.0018 nm for the CC and SP surface treatments, respectively.  Similarly, the bond length change between the Zn-O dimer one dimer into the bulk is 0.0063 nm and -0.0009 nm for the CC and SP surface treatments.  It is thus clear that while the surface atoms show different relaxations for the CC and SP surface treatments, in both cases there is a decrease in distance between the Zn-O dimer at the surface which is significantly larger than the distance change for the Zn-O dimer that is one dimer into the bulk, and furthermore the bond length decrease is much larger for CC than SP.  This relative decrease in surface dimer bond length explains the decrease in polarization in Fig. \ref{fig:ucthickness}, and thus effective piezoelectric constant in Table \ref{tab:eijeff}, and also why the decrease in effective piezoelectric constant in Table \ref{tab:eijeff} is much greater for the CC than SP surface treatments.  

To further investigate the validity of our calculated piezoelectric constants for the CC and SP surface treatments, we compare against existing DFT results for the effective piezoelectric constants of ZnO NWs.  Specifically, DFT calculations by~\cite{xiangAPL2006} and~\cite{Liangzhi:2008fn} also found a decrease in effective piezoelectric constants $e^{\text{eff}}_{\text{33}}$ and $e^{\text{eff}}_{\text{31}}$ with decreasing NW size, which is the same trend as found in the present work.  We note that the comparison is not exact, as the NWs in the DFT calculations had a hexagonal cross section oriented in the $[0001]$ direction that was enclosed by $(01\overline{1}0)$ surfaces.  However,~\cite{Liangzhi:2008fn} studied NW diameters from 3.1 to 0.4 nm and reported a decrease in $e^{\text{eff}}_{33}$ from 1.5 to 1.31 C/m$^{2}$, for a reduction of 14.5\%. In the present work, the reduction in the effective piezoelectric constant $e^{\text{eff}}_{\text{33}}$ of our original, CC and SP rectangular NWs with cross sectional sizes from 2-4 nm oriented along the $[0001]$ direction were found to be 14.9\%, 9.13\% and 13.0\%, respectively.

A final, but very important question to address is whether the CC and SP surface treatments will necessarily lead to a decrease in the effective piezoelectric constants of the NWs due to the fact that they reduce charge, and thus polarization at the NW surfaces.  While our results did in fact show a reduction in piezoelectric constant with decreasing NW size for all NW geometries and orientations considered, other literature results suggest that this need not be the case.  Specifically, we note the recent work of~\cite{agrawalNL2011}, who studied the piezoelectric properties of GaN and ZnO NWs, albeit with a hexagonal cross section as compared to the nearly square cross sections considered in the present work.  They also observed a charge and polarization reduction with decreasing NW size due to surface effects.  However, because of the strong contraction of the transverse surfaces, the reduction in volume (see Eq. (\ref{eqn:unicellP})), of the NW becomes more important than the reduction in surface charge, leading to a predicted increase in the effective piezoelectric constant for very small ($<$ 2 nm diameter) NWs.  These results also suggest that the cross sectional geometry may play a critical role in determining the size-dependence of the piezoelectric constant for NWs, as our preliminary studies also show an increase in effective piezoelectric constant with decreasing size for hexagonal ZnO NWs.

\section{Conclusions}

We have utilized classical molecular dynamics to study surface effects on the piezoelectric properties of ZnO nanowires with three different ($[2\overline{11}0]$, $[01\overline{1}0]$ and  $[0001]$) axial orientations.  A key finding is that treatment of the polar $(0001)$ surface via charge compensation or surface passivation is required to prevent the divergence of the electrostatic energy.  In the context of the atomistic simulations performed here, we demonstrated that not treating the surfaces to remove the electrostatic energy divergence results in spurious transformations of the initial wurtzite phase to a d-BCT phase.  With regards to the piezoelectric properties, the piezoelectric constants of the transformed d-BCT phase, which occurred for nanowires with untreated surfaces, were nearly one order of magnitude smaller than those calculated for nanowires whose surfaces had been treated using either the charge compensation or surface passivation techniques.  

Overall, our results show that the $[2\overline{1}\overline{1}0]$ oriented nanowires have a larger effective piezoelectric constant than the $[01\overline{1}0]$ oriented nanowires.  However, if proper treatment of the polar surfaces was performed, the effective piezoelectric constants for all nanowires were found to decrease with decreasing size, with all values smaller than the respective bulk ones.  We further demonstrated the underlying atomistic mechanism for the reduction in piezoelectric constants, in that regardless of whether the surface expanded or contracted in response to surface stresses, the bond length of the Zn-O dimer closest to the surface was found to decrease, thus causing a decrease in polarization at the nanowire surface and the corresponding reduction in effective piezoelectric constant.  Our overall finding is therefore that due to the observed decrease in piezoelectric constant for all three nanowire orientations with decreasing size, we recommend that larger diameter square or nearly square cross section nanowires be utilized in practical applications if maximum energy generation or harvesting using ZnO nanowires is desired. 

\section{Acknowledgements}

HSP and SD gratefully acknowledge support from the NSF, grant CMMI-0856261.

\bibliographystyle{plain}

% \bibliography{biball}

\end{document}